\documentclass[twocolumn,runningheads]{svjour2}
\smartqed  % flush right qed marks, e.g. at end of proof
\usepackage{graphicx}
\journalname{Astrophysics and Space Science}
\begin{document}

\title{Theoretical overview on high-energy emission in microquasars}
\author{Valent\'i Bosch-Ramon}
\institute{V. Bosch-Ramon \at
              Max Planck Institut f\"ur Kernphysik, Heidelberg 69117, Germany\\
              \email{vbosch@mpi-hd.mpg.de}}

\date{Received: date / Accepted: date}

\maketitle

\begin{abstract} Microquasar (MQ) jets are sites of particle acceleration and synchrotron emission. Such synchrotron radiation
has been detected coming from jet regions of different spatial scales, which for the instruments  at work nowadays appear as
compact radio cores, slightly resolved radio jets, or (very) extended structures  (e.g. \cite{mirabel99}, \cite{fender01},
\cite{corbel02}). Because of the presence of relativistic particles and dense  photon, magnetic and matter fields, these
outflows are also the best candidates to generate the very high-energy (VHE) gamma-rays detected coming from two of these
objects, LS~5039 and LS~I~+61~303 (\cite{aharonian05,aharonian06a}, and \cite{albert06}, respectively), and may be
contributing significantly to the X-rays emitted from the MQ core (e.g. \cite{markoff01}, \cite{bosch05a}).  In
addition, beside electromagnetic radiation, jets at different scales are producing some amount of leptonic and hadronic cosmic
rays (CR), and evidences of neutrino  production in these objects may be eventually found. In this work, we review on the
different physical processes that may be at work in or related to MQ jets. The jet regions capable to produce significant
amounts of emission at different wavelengths have been reduced to the jet base, the jet at scales of the order of the size of
the system orbital semi-major axis, the jet middle scales (the resolved radio jets),  and the jet termination point. 
The surroundings of the jet could be sites of multiwavelegnth emission as well, deserving also an insight. We focus
on those scenarios, either hadronic or leptonic, in which it seems more plausible to generate both photons from radio to VHE and
high-energy neutrinos. We briefly comment as well on the relevance of MQ as possible contributors to the galactic CR in the
GeV--PeV range. 
\end{abstract}

\section{Introduction}
\label{intro}

MQ are binary systems formed by a normal star and a compact object, either a black hole or a neutron star. In these objects,
stellar matter is expelled and some fraction of it eventually forms an accretion disk around the compact object from which two
relativistic jets are launched in opposite directions. These relativistic jets are the characterizing feature of MQ, namely a
subclass of X-ray binaries (XRB), and were observed for the first time in a galactic object at radio frequencies when studying
SS~433 (\cite{spencer79}, \cite{hjellming81}). In fact, the term {\it microquasar} became popular when   extended radio jets 
were discoverd in 1E~1740.7$-$2942 similar to those observed in some quasars (\cite{mirabel92}). We will focus on processes of
high energy emission in MQ, and for more general reviews we refer to other works that can be found in the literature (e.g.
\cite{mirabel99}, \cite{ribo02}, \cite{paredes05} and \cite{mirabel06}).

For several years after their discovery, MQ jets were considered basically as low energy emitters, producing
electromagnetic radiation at most up to X-rays (e.g. SS~433, \cite{kotani96}), and they were not generally
considered as gamma-ray emitters above $\sim 1$~MeV. There were some theoretical efforts to explain/predict
emission at higher energies from MQ
%\footnote{Gamma-ray models for XRB have been developed during decades (e.g.
%\cite{fabian77}, \cite{akharonian85}, \cite{bednarek90}, \cite{agaronian91}, \cite{dubus06a},
%\cite{khangulyan06}). Nevertheless, here we will focus on MQ models, in which jets or blobs are considered.} 
(see, e.g. \cite{atoyan99}), although at that time these objects were not conceived to be a class of gamma-ray
emitters of their own.

Paredes and co-workers (\cite{paredes00}) proposed the MQ LS~5039, discovered by the same authors, as the counterpart of the
high-energy gamma-ray source 3EG~J1824$-$1514, which had been detected by the high-energy gamma-ray instrument
EGRET (\cite{hartman99}). It was suggested in \cite{paredes00,paredes02} that inverse Compton (IC) interactions between
radio emitting relativistic electrons in the jet and UV photons from the massive stellar companion could be
behind the gamma-rays detected by EGRET, possibility that was further explored by \cite{bosch04a},
who concluded that it was feasible though synchrotron self-Compton (SSC) could not be discarded either. In fact,
the discovery of LS~5039 and its position coincidence with an EGRET source triggered an intense activity in the
field of MQ gamma-ray modeling (e.g. \cite{kaufman02}, \cite{romero03}, \cite{bosch05b},
\cite{dermer06}, \cite{paredes06}, etc.). The EGRET source association of this MQ was virtually
confirmed by the HESS detection of LS~5039 (\cite{aharonian05}), establishing it as the first TeV MQ. The
recent detection by MAGIC (\cite{albert06}) of another MQ, LS~I~+61~303 (\cite{massi01,massi04}), has shown
that these objects likely form a new population of gamma-ray sources. It is interesting to note that the TeV
emitter nature of LS~I~+61~303 renders very likely its association with the EGRET source 3EG~J0241$+$6103,
proposed by \cite{kniffen97}.

Beside gamma-ray photons, the production in jets of other high energy particles has been studied. Some MQ jets contain a
hadronic component, as has been observed in SS~433 (e.g. \cite{margon80}, \cite{kotani96}, \cite{marshall02},
\cite{migliari02}). Such hadronic component, in case it were a common MQ jet feature, would contribute to some extent to the
galactic CR in the GeV range (\cite{heinz02a}). In addition, if these hadrons were accelerated, they may affect significantly
the CR spectrum up to $\sim 100$~TeV (\cite{heinz02a}, \cite{bosch05c}, \cite{fender05}). Furthermore, the presence of 
relativistic protons in MQ jets, when interacting with dense matter and high energy photon fields, could lead to the production
of gamma-rays (\cite{romero03}, \cite{bosch05c}) and neutrinos (e.g. \cite{aharonian06b}, \cite{christiansen06}).

Next, in Sect.~\ref{scen}, the physical scenario is set up taking into account the present knowledge on MQ; in Sect.~\ref{reg},
different possible radiative processes occurring in MQ are briefly presented and discussed; finally, in Sect.~\ref{disc}, the
conclusions regarding the mechanisms explained here are provided and discussed in the context of the present observational
knowledge.

\section{The MQ physical scenario}
\label{scen}

The main components of the MQ scenario are the compact object, the stellar companion, the matter lost by the companion star,
the accretion disk, and the jet itself.  Here, the jet is considered as a conical region through which the emitting
relativistic particles flow confined inside. This region contains a magnetic field that is assumed to evolve attached to jet
matter, both being close to equipartition.  The matter lost by the star is embedding the jet and can form a very dense medium
for OB stellar companions up to relatively large distances. The star in these massive systems is as well a bright source of
UV photons and  generates a powerful magnetic field (a characterization of the magnetic field of a massive star can be found
in, e.g., \cite{usov92}).  Since massive systems present some features, lacking in low-mass MQ, that introduce further
complexity to the overall emitting scenario, we will attach ourselves to the high-mass MQ case, although many of the
discussed processes could take place as well in low-mass objects. 

\subsection{Studying high energy emission in the MQ scenario}

Using the prescription for the MQ scenario given above, some estimations are made and presented in this work for the spectral
energy distribution (SED) of the radiation that could be produced in MQ different regions. In our approach the jet plays a very
important role. Each emitting zone is treated as homogeneous, with a size being similar to the corresponding typical spatial
scale. Bremsstrahlung, synchrotron and (Thomson and Klein-Nishina) IC losses are taken into account for electrons, and
convective losses are considered for both electrons and protons, the latter being not affected by radiative losses in most of
cases. To set up the jet properties involved in the performed computations, we have adopted the next values of the relevant
parameters of the MQ system: a jet Lorentz factor of 1.1; accretion disk luminosity of $10^{36}$~erg~s$^{-1}$; disk photon
energy of 10~keV\footnote{It is common to assume the existence of a hot cloud in the innermost accretion disk region so-called
corona that would present a different spectrum than the accretion disk (see, e.g. \cite{esin97}). For simplicity, we consider
the photon fields generated in the disk and the corona as the same one, with a mean photon energy between those of the disk and
corona photons.}; a stellar companion luminosity of $10^{39}$~erg~s$^{-1}$; stellar photon energies of 10~eV; an orbital
semi-major axis of 0.2~AU; a stellar wind mass-loss rate of $10^{-6}$~M$_{\odot}$~yr$^{-1}$; and a stellar wind velocity of
$2\times 10^8$~cm~s$^{-1}$.  The jet and the wind velocities can be taken as the convection velocities in the jet and the wind,
respectively, which altogether with the region size define the convection timescales. For the jet termination region, a
convection velocity of $\sim 3\times 10^9$~cm~s$^{-1}$ has been adopted assuming downstream jet material convects away at the
speed of the postshock region. All these physical quantities can be in principle known from observations, and the values
adopted here are typical ones from the literature. All the parameters values are given in Table~\ref{tab:1}, at the top.

There is a small set of parameters which are free because they refer to physical quantities that are not well constrained  at
the moment. It is taken a jet kinetic luminosity of $10^{36}$~erg~s$^{-1}$; a injection luminosity of the relativistic particles
of $10^{35}$~erg~s$^{-1}$; a magnetic field ($B$) of 1/10 the equipartition value along jet, the equipartition value for the
termination region and 1~G for the jet surrounding region; a mean jet particle energy dominated by the bulk motion component
(i.e. cold-matter dominated jet) even when there is relativistic particle component (concerning the jet average temperature, see
below); a minimum Lorentz factor of relativistic electrons of 100; a minimum Lorentz factor of relativistic protons of 1; and an
acceleration efficiency (times $qBc$) of 0.1. The power-law index of the particle injection spectrum is fixed to 2. For the
interstellar medium density, a value of 10$^3$~cm$^{-3}$ is taken. All these free parameter values have been adopted similar to
those typically found in the literature and have been summarized in Table~\ref{tab:1}, at the bottom.

The highest energy particle is derived from the acceleration efficiencies in balance with the radiative losses and the
accelerator size, leading to GeV energies for electrons in the jet innermost regions and $>$TeV elsewhere; protons are not
cooling restricted, being able to reach $>$TeV everywhere in the jet (see e.g. \cite{rieger06}). The adopted acceleration
efficiency allows for instance to reach TeV photon production energies for the emitting particles. The minimum electron
Lorentz factor is derived from the thermal energy of protons, assuming a proton temperature of $10^{11}$~K. The magnetic
field has been fixed to avoid too strong synchrotron radiation, although its choice is somewhat arbitrary. All this allows to
estimate roughly the photon and neutrino production all along the jet. {\bf The emission generated by the secondary pairs
produced via charged pion and then muon decay is not computed due to the low luminosities of the neutral pion decay gamma
component, rendering the secondary radiation negligible when compared with the one of the primary leptonic component.}
Further issues related to the exact spectral shape and its evolution: {\bf photon photon absorption} and cascading,
variability in target densities or particle injection, geometric effects in the different interaction processes (IC and
photon photon absorption), and Doppler boosting in relation to the observer direction, are not studied quantitatively here
and deserve a particular treatment source by source out of the scope of this work. {\bf SSC is not taken into account either,
although it will not affect significantly the electron energy distribution but for the case when the emitter is very compact
(i.e. the base of the jet). In such a case, the synchrotron component would be weaker though not much, and the SSC would be
similar in flux, although broader in spectral shape, than the disk IC component (see next section).}

{\bf It is worth before going further to briefly mention the main difference between a low-mass and a high-mass system, the
one treated here, concerning gamma-ray emission. As will be seen, the stellar IC is an efficient mechanism to produce
significant amounts of TeV photons, implying that low-mass systems lacking of a strong source of optical/UV photons may not be
strong emitters.}

\begin{table}[t]
\caption{Parameter values adopted in this work}
\centering
\label{tab:1}       
\begin{tabular}{ll}
\hline\noalign{\smallskip}
Parameter & Value  \\[3pt]
\tableheadseprule\noalign{\smallskip}
Jet Lorentz factor & 1.1 \\
Accretion disk luminosity & $10^{36}$~erg~s$^{-1}$ \\
Disk photon energy & 10~keV \\
Stellar companion luminosity & $10^{39}$~erg~s$^{-1}$ \\
Stellar photon energy & 10~eV \\
Orbital semi-major axis & 0.2~AU \\
Stellar mass-loss rate & $10^{-6}$~M$_{\odot}$~yr$^{-1}$ \\
Stellar wind velocity & $2\times 10^8$~cm~s$^{-1}$ \\
Jet-ISM postshock speed & $3\times 10^9$~cm~s$^{-1}$ \\
\tableheadseprule\noalign{\smallskip}
Jet kinetic luminosity & $10^{36}$~erg~s$^{-1}$ \\
Particle injection luminosity & $10^{35}$~erg~s$^{-1}$ \\
Ratio magnetic field/equipartition value & 0.1 \\
Electron minimum Lorentz factor & 100 \\
Proton minimum Lorentz factor & 1 \\
Acceleration efficiency & 0.1 \\
ISM density & 10$^3$~cm$^{-3}$ \\
\noalign{\smallskip}\hline
\end{tabular}
\end{table}

\subsection{Further comments on the jet properties}

Relativistic particles plus magnetic field are unlikely to be the only components of the jet. Relativistic particles lose their
energy as far as they are convected away in the jet under the present magnetic and photon fields, and they have to be accelerated
far away from the compact object when extended radio/X-ray emission is observed. Moreover, emitting particles must be confined
to keep the (supposed to be) conical jet shape with a small opening angle (e.g. SS~433, \cite{marshall02}; LS~5039,
\cite{paredes02}). It implies the existence of a non-emitting background outflow that dominates dynamically the whole jet,
sometimes called {\it dark jet}. We are not concerned on this background here since its detailed physics is not needed for our
exploration. The background jet action on the radiative part is taken into account phenomenologically, assuming this background
confines and accelerates particles, which is enough at the level of approximation of our exploration. 

The dark jet might consist on matter extracted from the inner parts of the accretion disk (strongly ionized matter with
temperatures $\sim 10^7$~K) and/or the corona (a hot matter halo with temperatures of $\sim 10^{11}$~K for protons and $\sim
10^9$~K for electrons around the compact object; e.g. \cite{narayan95}). The mean energy per particle of this matter,
redirected from the accretion disk/corona to the jet, would be non relativistic (i.e. cold), kinetically confining the jet due
to its low Mach number. This cold matter would prevent efficient relativistic particles to escape via a low enough diffusion
coefficient. Another possible type of dark jet would be that formed by a light and relativistic pair beam that would be
(re-)energized by magnetic energy dissipation due to instabilities occurring in a surrounding magnetically dominated flow. Such
flow would be dynamically dominant and would play the role of the jet confining component. Each case would be associated to
different launching mechanisms, accretion vs. black hole rotation (e.g. \cite{blandford82}, \cite{blandford77}, respectively),
although a combination of all these possibilities could be the case (for recent proposals for jet launching, see e.g.
\cite{meier03}, \cite{hujeirat04}, \cite{ferreira06}). Other on-going debates concern the discrete or continuous nature of the
outflows ejected from MQ, and which radio/X-ray states different types of jet are related to. For a recent treatment of the
problem, we refer to \cite{fender04}. Due to the simplicity of our approach, we consider the jet as a continuous structure in
which acceleration and emission processes can take place at different locations. Nevertheless, it is worth noting that the
outflow could suffer strongly time-dependent injection, it could be formed by discrete ejections, or it may be a chain of
active knots on a background non emitting continuous flow. Such knots could be produced by either internal shocks between
different velocity shells, dissipation of energy via magnetic instabilities, or fast Alfven waves moving fast through the
background matter (maybe all of them altogether). 

\section{The MQ regions}\label{reg}

\begin{figure}
\centering
  \includegraphics[width=0.38\textwidth]{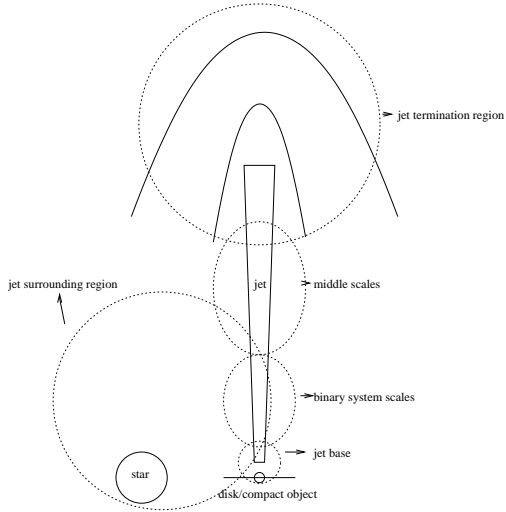}
\caption{Sketch of the MQ five considered regions.}
\label{fig:1}       
\end{figure}

A MQ can be divided in five regions according to the dominant conditions that affect the radiative processes: the jet base, at
100 Schwarzschild radii from the compact object ($\sim 100 R_{\rm Sch}\sim 10^8$~cm, for a compact object of $\sim
3$~M$_{\odot}$); the jet at distances similar to the binary system size, $\sim 3\times 10^6$~$R_{\rm Sch}$\footnote{This region
would be imposed by the presence of a massive star in the system since in low-mass MQ it is unclear whether the star could be
something else than a matter supplier, without effects in the radiation processes.}; the middle scales, at distances $\sim
10^9$~$R_{\rm Sch}$; the jet termination region, at $\sim  10^{12}$~$R_{\rm Sch}$; and the regions surrounding the jet, with
typical sizes that would depend on the jet environment (i.e. the stellar wind in a massive system). For the jet surrounding
region, the size of the emitter is taken that of the binary system. For the remaining regions, the jet size will be taken one
fifth of the distance $l$, which may appear somewhat arbitrary but its precise value does not affect significantly our
estimates. In Fig.~\ref{fig:1}, the five regions are shown. In Table~\ref{tab:2} we present the sizes and
distances to the compact object corresponding to the different mentioned regions.

\begin{table}[t]
\caption{The MQ regions}
\centering
\label{tab:2}       
\begin{tabular}{lll}
\hline\noalign{\smallskip}
Region & distance & size  \\[3pt]
\tableheadseprule\noalign{\smallskip}
Jet base & 10$^8$~cm & $2\times 10^7$~cm \\
Binary system scales & $3\times$10$^{12}$~cm & $6\times 10^{11}$~cm \\
Middle scales &  10$^{15}$~cm & $2\times 10^{14}$~cm\\
Jet termination point &  10$^{18}$~cm & $2\times 10^{17}$~cm \\
Jet surroundings &  $3\times 10^{12}$~cm & $3\times 10^{12}$~cm\\
\noalign{\smallskip}\hline
\end{tabular}
\end{table}

\subsection{The jet base}

The innermost jet is usually associated to the inner accretion disk, where the jet is formed. This would
correspond to a region in AGN jets of $\sim 10^{16}$~cm, which is observable in some nearby objects, like M~87 or
Centaurus~A (\cite{junor99} and \cite{horiuchi06}, respectively). Unfortunately, unlike in AGN, the jet
base size in MQ is several orders of magnitude away from the present resolution capabilities of the radio
interferometers at work. The linear size is proportional to the compact object mass, but the angular size is
inversely proportional to the distance, leading to a resolution capability $10^5$ times better for the nearest
AGN than in case of MQ, preventing to probe this region of the jet for the latter objects. 

To compute the emission from leptons and hadrons in the jet base, there are several elements to take into account. These
elements are the magnetic field, the accretion disk photon field and the jet hadrons\footnote{These hadrons may be also disk or
corona ions, if mixing with jet relativistic protons takes place.}, the last two being much denser there than the stellar photon
and matter fields. The mechanisms behind particle acceleration in this region are unclear. If the jet base were magnetically
dominated (as considered by e.g. \cite{sikora05}), one of these mechanisms could be magnetic energy dissipation via MHD
instabilities, which would accelerate particles (e.g. \cite{zenitani01}). Also, if jet velocities were high enough in the jet
base, the dense available photon and matter fields could allow the converter mechanism to take place (\cite{derishev03}).
Reconnection of large scale magnetic fields in the surrounding corona could inject a non-thermal population of particles in the
jet as well (see, e.g., \cite{gierlinski03} and references therein).

\begin{figure}
\centering
  \includegraphics[width=0.38\textwidth]{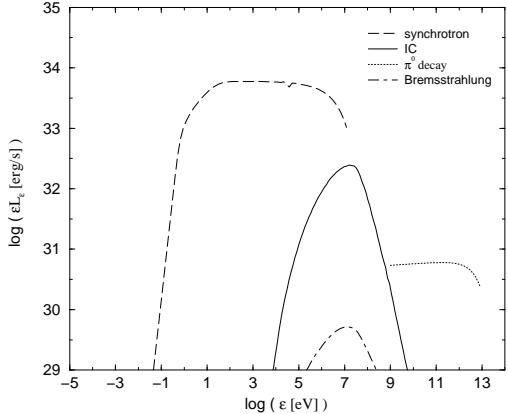}
\caption{SED for the synchrotron and IC processes, (proton-proton) $\pi^0$ decay, 
and Bremsstrahlung in the jet base under the conditions given in Sect.~\ref{scen}. We note that photon absorption 
is not considered and only the production SED is presented.}
\label{fig:2}       
\end{figure}

Depending on the dominant conditions, the relevant leptonic radiative mechanisms in the jet base could be synchrotron
emission (e.g. \cite{markoff01}), Bremsstrahlung by interacting with jet ions (\cite{bosch06a}), SSC (e.g.
\cite{bosch04a,bosch04b}) and IC with corona and/or disk photons (e.g. \cite{romero02}). Regarding hadronic processes, the
radiative mechanisms that could produce gamma-rays, neutrinos and, as a by-product, low energy emission from secondary pairs
are the interactions of relativistic protons with ions in the jet, and interactions between jet relativistic protons and
X-ray photons from the disk, the corona or the jet itself (e.g. \cite{levinson01}, \cite{aharonian06b}). These proton ion
collisions would produce neutral pions ($\pi^0$) that decay to gamma-rays, and charged pions ($\pi^{\pm}$) that decay to
muons and neutrinos, the former decaying then to electron-positron pairs and neutrinos. The presence of strong X-ray photon
fields renders the possibility of $\le$~GeV gamma-ray absorption cascading quite likely (e.g. \cite{akharonian85},
\cite{wu93}). In Fig.~\ref{fig:2}, we show the SED for the synchrotron, IC, (proton-proton) $\pi^0$ decay and Bremsstrahlung
components computed for the model described in Sect.~\ref{scen}. It is seen that synchrotron is the dominant radiative
mechanism, with strongly self-absorbed radio emission, partially cooled spectrum at high energies, and maximum photon
energies reaching hard X-rays. We do not compute neutrino emission, although for the cases when the proton proton gamma-ray
component is shown, the corresponding neutrino SED is to be similar (neglecting absorption). 

The variability of the radiation coming from the jet base could be very fast due to the involved spatial scales and is beyond
the timing capabilities of the present gamma-ray instruments.

\subsection{The binary system scales}

At the binary system scales, the jet in radio appears still as core emission, although the corresponding size in an AGN jet
would be tens of pc and can be resolved by the present interferometers. In high-mass systems, the magnetic field in the jet, if
in equipartition, competes with the stellar photon field concerning lepton cooling. Moreover, convective and adiabatic losses
are as well significant for the lepton spectrum at lower energies. The stellar wind at these scales is perhaps a good target for
jet proton wind ion collisions, as well as for jet electrons, if wind/jet matter mixing is significant. Plausible mechanisms for
these scales to generate relativistic particles are the different versions of the Fermi process: shock diffusive (Fermi I),
random scattering (Fermi II) and shear acceleration (e.g. \cite{drury83}, \cite{fermi49} and \cite{rieger04}, respectively; see
also \cite{rieger06}). Fermi I mechanism could take place due to internal shocks in the jet; Fermi II acceleration could take
place if magnetic turbulence is significant enough; shear layer would be a natural outcome of an expanding jet or different
jet/medium velocities. Interaction with
the stellar wind may also trigger particle acceleration.

\begin{figure}
\centering
  \includegraphics[width=0.38\textwidth]{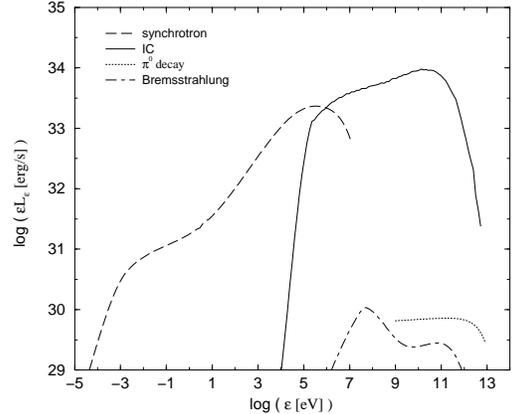}
\caption{The same as in Fig.~\ref{fig:2} but for the jet emission at binary system scales.}
\label{fig:3}       
\end{figure}

From the different elements playing a role at these scales within the microquasar context, possible significant emitting leptonic
processes could be synchrotron emission (e.g. \cite{yuan05}, \cite{bosch06a}, \cite{paredes06}), Bremsstrahlung (e.g. \cite{bosch06a}),
and IC (e.g. \cite{paredes00,paredes02}, \cite{kaufman02}, \cite{georganopoulos02}, \cite{bosch04a}, \cite{dermer06}{\bf,
\cite{gupta}}). Hadronic processes that may play a significant role at binary system scales could be proton proton collisions with the
stellar wind (e.g. \cite{romero03}), which would lead as well to neutrino production (e.g. \cite{romero05}, \cite{aharonian06b}). In
Fig.~\ref{fig:3}, we show the SED for the synchrotron, IC, $\pi^0$ decay and Bremsstrahlung components produced in the jet at binary
system scales. We note that stellar IC and synchrotron processes compete at some extent being both significant. Synchrotron emission is
still quite self-absorbed at radio frequencies, and compete with IC at X-rays, which reaches VHE. We recall that photon photon
absorption (\cite{bottcher}, \cite{dubus06b}) and electromagnetic cascading (e.g. \cite{aharonian06b}, \cite{romero06}) are not
computed and only the production spectra are shown.

Variations of high energy emission coming from these jet scales could be detected, and could be linked to orbital
changes of the emission and absorption processes (injection of relativistic particles, changes in targets, etc.).

\subsection{The jet middle scales}

MQ jet middle scales are already within the resolution capabilities of the present interferometers though extended radio
emission is generally faint. These regions would have a size of $\sim 10$~kpc in AGN. In this region of the MQ jet, the
outflowing matter is probably overpressing the surrounding medium\footnote{Such environment is assumed to be similar to the
typical galactic ISM, although it could be much more complicated in high-mass systems, since the stellar wind cannot be
neglected.} and in such a case no significant interaction will be expected, unlike in extragalactic jets. Regarding particle
acceleration, although internal shocks may take place still at these distances from the compact object associated perhaps with
violent episodes occurring in the jet base, Fermi II type and shear acceleration appear more plausible for a continuous outflow
at these scales (something similar could happen in the intra-knot regions of extragalactic jets, see, e.g., \cite{rieger06}).

\begin{figure} 
\centering 
\includegraphics[width=0.38\textwidth]{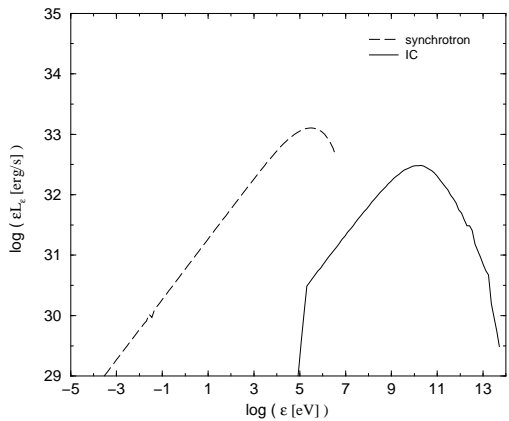} 
\caption{The SED for the synchrotron and IC components for the jet emission at middle scales.} 
\label{fig:4}        
\end{figure}

At the scales concerned here, only synchrotron emission could be significant in persistent jets, and stellar IC seems already a
bit inefficient (e.g. \cite{bosch06a}) and could be masked by emission coming from inner regions. The particle distribution
evolution is likely dominated by convective and adiabatic losses (\cite{vanderlaan66}). Nevertheless, SSC could still be
significant for powerful ejections (e.g. \cite{atoyan99}). In Fig.~\ref{fig:4}, we present the computed SED for the synchrotron
and IC components of the radiation produced at jet middle scales. Neither Bremsstrahlung nor proton proton collisions are likely
to be relevant for the expected low densities of the jet and its environment. Using the parameters provided above, it is seen
that synchrotron radio emission is faint and optically thin and dominant losses are convective ones. X-ray emission could be
still significant.

The variability of the radiation at this scales could come from long term changes in the system properties, like
for instance the mass loss rate of the star.

\subsection{The jet termination region}

At the jet termination point, as in AGN hot spots and radio lobes, two shocks may be formed or become important when the
swept ISM inertia starts to affect the jet advance, one moving backwards in the jet (i.e. the reverse shock), and another one
moving forward (i.e. forward shock). Fermi I type mechanism seems the most reasonable option under these conditions, although
the diffusive and convective efficiency in the downstream regions of the forward/reverse shocks could prevent efficient
acceleration to occur. For particles accelerated in such shocks, synchrotron, Bremsstrahlung and IC processes in case of
leptons and proton proton collisions in case of hadrons could produce non negligible amounts of radiation. It might be the
case as well that hydrodynamical instabilities distorted the jet and mixed jet matter with the ISM without forming a strong
shock (for a thorough discussion on this region properties, see \cite{heinz02a}). We note that even without shock, some
interaction may still take place between jet hadrons and ISM nuclei.

It is difficult to ascertain whether the jet termination region can accelerate particles efficiently.  Extended X-ray
synchrotron emission has been observed from the MQ XTE~J1550$-$564 likely produced by accelerated electrons in a termination
jet shock (\cite{corbel02}, see also \cite{wang03}). Otherwise, extended radio emission at pc scales has been detected, for
instance, coming from 1E~1740.7$-$2942, which could be related to the interaction of the MQ jets with a surrounding molecular
cloud (\cite{mirabel92}). Other examples of the interaction between a galactic jet and its environment are SS~433 and
Cygnus~X-1 (\cite{dubner98} and \cite{gallo05}, respectively). Therefore, although a different morphology has been found for
the jet largest scales in AGN and MQ, it could be explained by the differences in scaling of the physical quantities involved
in the jet-medium interaction process in galactic and extragalactic jets (e.g. \cite{heinz02b}), i.e. one cannot simply scale
with for instance the black hole mass, since the environment does not need to be correlated with it (at least in a
straightforward manner). 

\begin{figure}
\centering
  \includegraphics[width=0.38\textwidth]{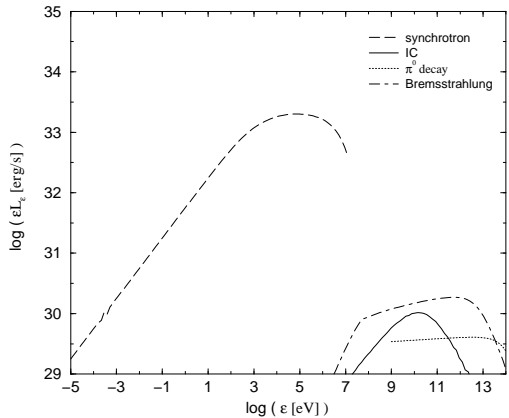}
\caption{The same as in Fig.~\ref{fig:2} but for the jet termination point radiation.}
\label{fig:5}       
\end{figure}

In case particle acceleration and confinement would be efficient, synchrotron radiation could be produced presenting moderate
fluxes at few kpc distances (e.g. \cite{bosch06b}). Hadronic acceleration could take place as well, which could lead to
gamma-ray production (e.g. \cite{heinz02a}) and secondary leptonic emission if embedded in or close to a dense medium (e.g.
\cite{bosch05c}). It seems that such relativistic hadrons would hardly dominate the CR galactic component in our galaxy (e.g.
\cite{fender05}), although the cold part could still be significant (e.g. \cite{heinz02a}). In Fig.~\ref{fig:5}, we show the SED
of the synchrotron, IC, $\pi^0$ decay and Bremsstrahlung components for the parameters described above. In the presented
situation, synchrotron emission is significant at all wavelengths from radio to hard X-rays, and the other mechanisms are
negligible\footnote{It does not imply that it should be the case in general. For high hadronic densities, or for lower
convective velocities of the shocked ISM, it could be possible to get higher gamma-ray fluxes. The parameter setting is probably
less constrained for this case than previous ones because observational data is sparse.}.

The timescales at which radiation can change where the jet terminates are large, of at least several years, being
difficult to follow.

\subsection{The jet surroundings}

\begin{figure}
\centering
  \includegraphics[width=0.38\textwidth]{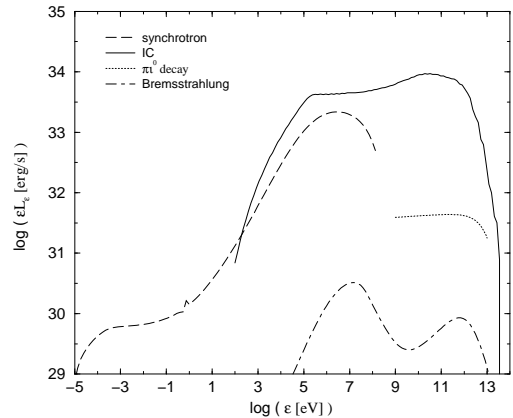}
\caption{The same as in Fig.~\ref{fig:2} but for the jet surroundings.}
\label{fig:6}       
\end{figure}

Jets are likely efficient confining relativistic particles. Typically, accelerated particles are considered to reach a maximum
energy in jets that, when cooling is inefficient enough, corresponds to a typical diffusion length equal to the size of the
jet. Actually, it does not need to be the case all along the jet. Confinement could fail because local changes of the jet
conditions, mixing with the surrounding medium, etc. Moreover, for systems in which photon photon opacities are high for a
significant fraction of the whole solid angle, as seen from the gamma-ray emitter, great amounts of pairs will be created when
gamma-rays interact with the ambient photons outside the jet. All this may lead to the injection of a relativistic population
of particles in the jet surroundings, likely with high minimum energies. Perhaps the most interesting case to study is that
when particles are injected at binary system scales in a massive system, since particles could be confined within (as secondary
particles generated by cascading; see, e.g., \cite{bednarek06}). Confined particles would move with the wind allowing for more
efficient energy dissipation of this particles, since the convection velocity of the wind is almost two orders of magnitude
lower than that of the jet. For leptons, there are significant matter, magnetic and photon fields to interact with, which may
lead to significant synchrotron and IC radiation. For hadrons, the wind ions represent a dense target for hadronic interactions
which could lead to neutrino and gamma-ray production (e.g. \cite{bednarek05}, \cite{aharonian06b}). In Fig.~\ref{fig:6}, we
show the SED synchrotron, IC, $\pi^0$ decay and Bremsstrahlung components in the jet surroundings using the parameters given in
Sect.~\ref{scen}. In the present scenario, synchrotron emission is significant at X-rays, where competes with IC emission, and
it is non negligible either in the radio band though suffer strong self-absorption (and may suffer absorption in the stellar
wind) at centimetric wavelengths. IC radiation is also a very efficient process producing gamma-rays, that should be affected
as well by photon photon absorption and cascading, which is not considered here.

Regarding variability, it will depend on the energy of particles and the properties of the medium where these particles will be
injected. In a high-mass system, orbital variability of the radiation coming from the most energetic particles could take place
due to injection changes along the orbit. Low energy particles would present much less variability.

\section{Discussion}
\label{disc}

Taking into account all the possible processes described above, let us look at the observational results to compare them with
the expectations from the different theoretical models. It is worth noting that it has not been possible till recently, less
than one decade ago, to probe high energy radiative processes in MQ because of the lack of data in that energy domain. There
were some claims of detection of Cygnus~X-3 at gamma-rays by instruments working at VHE and UHE instruments (e.g.
\cite{chadwick85}), although these detections are considered doubtful nowadays by the VHE community. In any case, a
significant production of gamma-rays above 1~MeV in MQ was not thought as plausible. At that time, the main practical reasons
behind the difficulties detecting microquasars were that, on one hand, sensitivity was not high enough to reach reliable
detections and, on the other, poor angular resolution rendered identification of the counterpart at lower energies a very
tough task. However, since the proposal of LS~5039 as an EGRET source (\cite{paredes00}), MQ scenario started to be explored
more actively from the theoretical and the observational points of view by groups coming from different fields (AGN, GRB,
pulsars, radio stars, XRB, etc.). Nowadays, there are evidences brought by the new instrumentation supporting MQ as a
population of gamma-ray sources. The facts behind those evidences can be linked with the described mechanisms that could be
at work in MQ. 

{\bf Large scale jets}

It is well known, as has been noted above, that MQ jets contain relativistic particles emitting synchrotron radio emission at
spatial scales that comprehend from those not resolved by the present interferometers to the termination jet region. Moreover,
some MQ have shown non-thermal X-ray emission from large-scale jets, which is evidence of the presence of TeV particles and
external shocks (e.g. \cite{corbel02}). Thus, all this tells us that leptonic particles can reach TeV energies in the jet
termination regions. Such TeV leptons can produce IC TeV radiation if a significant source of photons is nearby. In addition,
ions are known to be heated in jets up to quite large distances (e.g. \cite{migliari02}), and could be the energy carriers at
least in some MQ jets (e.g. \cite{gallo05}). This suggests the possibility of hadronic acceleration and $\pi^0$-decay gamma-ray
and neutrino production by proton proton collisions, with lower energy photons generated via synchrotron emission from the
$\pi^{\pm}$-decay pairs as by-products. The current upper-limits imposed by HESS and MAGIC on the extended TeV emission from MQ
make the detection of extended neutrino emission quite unlikely for the near future (at such spatial scales, photon photon
absorption cannot mask the overall flux of VHE photons, being similar thus to the neutrino one), but extended TeV emission may
still be detected in the near future either by HESS and MAGIC or by HESS- and MAGIC-II.

{\bf Variable TeV emission}

Two MQ have been clearly detected at TeV energies: LS~5039 and LS~I~+61~303 (\cite{aharonian05} and \cite{albert06},
respectively); both objects also proposed to be EGRET sources (\cite{paredes00} and \cite{kniffen97}, respectively), which has
been virtually confirmed by their VHE detections. Concerning the physical processes occurring in microquasars, much information
can be obtained from this recent VHE data of MQ. TeV emission from LS~I~+61~303 has shown to be variable at scales similar to
those of the orbital period, pointing to a strong link between the emission processes and the change of the emitter properties
along the orbit (\cite{albert06}). For LS~5039, strong TeV variability, orbital periodicity and absorption effects on the VHE
radiation  shows that this emission is likely coming from a small region $\le$ few AU (\cite{aharonian06a}). Therefore,
hadronic and/or leptonic acceleration is probably taking place at binary system scales. In the context of leptonic models, the
strong losses and high photon photon internal opacities in the base of the jet (e.g. \cite{bosch06a}) will likely prevent that
region from being a significant TeV emitter, pointing to stellar IC and the jet at binary system scales as the radiative
process and the emitting region, respectively, behind the VHE radiation. For hadronic processes, radiative losses are not
relevant enough in general to limit proton energy, although requiring as well very large target densities and kinetic
luminosities in relativistic hadrons. Cascading seems unavoidable for a jet base hadronic model. The higher efficiency of IC
interaction leaves more room for significant TeV photon production outside the binary system, although still close to it ($\sim
1$~AU).

The details concerning the injection particle distribution, or at which extent photon photon opacities affect the final TeV
spectrum, are still unknown. Different particle energy distributions, with different evolution and overall shape because
different injection location and jet conditions, may be contributing to the final gamma-ray spectrum. Because of the emitting
size is not properly settled, it seems not possible at the moment to infer the local production spectrum applying only an
absorption or a cascading model. Due to the complexity of leptonic and hadronic mechanisms, it is hardly feasible presently to
put forward a method for favoring with strong evidence any of both scenarios but via neutrino detection. Nevertheless, the
present detected luminosities in the TeV regime show that, unless these systems produce all the emission under high photon
photon opacities rendering a high neutrino to photon ratio, it will be difficult to detect them even with $km^3$ neutrino
detectors (e.g. \cite{aharonian06b}, \cite{christiansen06}, \cite{torres06}, \cite{bottcher2}).

One interesting fact concerning both the X-rays and TeV gamma-rays is that they appear to be correlated in LS~5039. There is a
clear peak and a spectral hardening at the same orbital phase, $\sim 0.8$, in both X-rays and TeV energies (\cite{bosch05a},
\cite{casares05}, \cite{aharonian06a}). This could happen as well in LS~I~+61~303, where the X-ray and the TeV peaks occur at
similar phases $\sim 0.5$ (\cite{taylor96}, \cite{albert06}), and there is also a hardening in the X-ray spectrum when X-ray
flux increases (\cite{sidoli06}). There might be also a strong connection between radio and TeV photons in these two TeV MQ. In
both systems, most of the radio emission come from scales of several AU, but it is optically thin (in LS~I~+61~303 when it is
not in outburst; \cite{ray97}), which points to moderate magnetic fields and particle densities. Moreover, LS~5039 shows quite
steady radio fluxes (\cite{ribo99}). This former fact, altogether with the compactness of most of the radio emission and the
mentioned moderate magnetic field, would imply that the radio emission is to be confined (e.g. a low convective velocity, like
that of the stellar wind and unlike the jet one). We note that TeV emission, likely being of stellar IC origin, requires
magnetic fields not far from those of radio emission ($\le 1$~G, provided the stellar photon field density is $\sim
500$~erg~cm$^{-3}$), which may mean that radio and TeV radiation come from similar regions. However, since TeV luminosities are
slightly smaller than the X-ray luminosity upper-limit and TeV spectrum is not affected by dominant synchrotron losses, the
hypothesis that the same population of electrons produces both X-ray and TeV photons seems to be ruled out. An interesting
additional fact is the constant 200~GeV flux in LS~5039 (\cite{aharonian06a}), which may be related to the radio flux
steadiness.

{\bf Multiparticle future}

Previous discussion shows that a deep understanding of the physics behind the high energy radiation observed in MQ requires of
the combination of different processes taking into account particle acceleration, radiation mechanisms, photon photon absorption
and cascading, and plasma physics. Nevertheless, this approximation to the problem cannot be done at once, and a step by step
study is to be applied accounting for all the existent (and forthcoming) data in combination with theoretical insights on the
relevant processes and emitting regions. It is clear that multiwavelength and multiparticle approaches, plus deeper modeling,
are indeed the future of the MQ high energy field. 

\begin{acknowledgements}
%We thank an anonymous referee for useful comments and suggestions. 
We are grateful to Josep M. Paredes and Marc
Rib\'o for a thorough reading of the manuscript. We thank also Felix Aharonian, Dmitry
Khangulyan and Evgeny Derishev for fruitful discussion that has benefited this work. 
V.B-R. acknowledges partial support by DGI of the Ministerio de Educaci\'on y Ciencia (Spain) under grant
AYA-2004-07171-C02-01. The Max-Planck-Institut f\"ur Kernphysik is thanked for its support and kind hospitality. 
\end{acknowledgements}

\end{document}